\documentstyle[aps,prl,twocolumn]{revtex}

\def\ot{\otimes}
\def\ra{\rangle}
\def\la{\langle}
\def\id{i\!d}
\def\ie{{\em i.e.\ }}

\def\1{\uparrow}
\def\0{\downarrow}
\def\s{\sigma}
\def\su2{\mbox{SU(2)}}
\def\suq2{\mbox{SU${}_q$(2)}}
\def\xp{X^+}
\def\xm{X^-}
\newcommand{\beq}{\begin{equation}}
\newcommand{\eeq}{\end{equation}}
\newcommand{\bea}{\begin{eqnarray}}
\newcommand{\eea}{\end{eqnarray}}

\newcommand{\zet}{\hbox{\bf Z}}

\newcommand{\hhub}{H_{\mbox{\scriptsize Hub}}}
\newcommand{\hloc}{H^{\mbox{\scriptsize (loc)}}}

\newcommand{\hsym}{H_{\mbox{\scriptsize sym}}}
\newcommand{\hqsym}{H_{\mbox{\scriptsize $q$-sym}}}

\newcommand{\hext}{H_{\mbox{\scriptsize ext}}}

\begin{document}
\title{
{\rm\mbox{}\hfill LMU-TPW 1996-29 and LBNL-39570\\[1em]}
{\Large\bf Quantum Symmetry of Hubbard Model Unraveled}}
\author{Peter Schupp\\{\sl Sektion Physik, Universit\"at M\"unchen\\
Theresienstra\ss e 37, D-80333 M\"unchen\\
Federal Republic of Germany\\
{\small peter.schupp@physik.uni-muenchen.de}}\\[1em]
\parbox{16cm}{\small Superconducting quantum 
symmetries in extended one-band one-dimensional Hubbard models
are shown to  originate from the classical (pseudo-)spin symmetry
of a new class of models; the standard
Hubbard model is a special case. 
The quantum symmetric models provide extra parameters but are 
restricted to one dimension.
All models discussed 
are related by generalized Lang-Firsov transformations, some
have symmetries away from half filling.\\[1ex]
\vspace{1em}}
}
\date{October 1996}
\maketitle

The exploration of high temperature superconductivity in cuprates 
has greatly revived interest in the Hubbard model \cite{H} as a model of
strongly correlated electron systems \cite{A,AS,Da}.
Despite its formal simplicity this model continues to resist complete
analytical or numerical understanding.  
Symmetries of the Hubbard Hamiltonian play a major role in the 
reduction of the problem. They have for instance been used  to construct eigenstates of the Hamiltonian with off-diagonal long range order 
\cite{Y}, to simplify numerical diagonalization\cite{BA} and
to show completeness of the solution \cite{LW} to
the one-dimensional model \cite{EKS}.
Since the work of Heilmann and Lieb \cite{HL} the Hubbard model
is known to have coupling constant independent symmetries (spin and
pseudospin) and coupling constant dependent ``hidden'' symmetries in
addition to spatial symmetries for special configurations
(benzene-hexagon in \cite{HL}). Operators corresponding to
the hidden symmetries were found in \cite{S,G}; see also the
nice review \cite{L}. 

The (pseudo-)spin SO(4) symmetry \cite{HL,Y,YZ}
of the standard Hubbard model is restricted to
the case of an average of one electron per site  
(half-filling), so recent speculations \cite{MR} about 
extended Hubbard models with coupling constant independent generalized
(quantum group) symmetries away from half-filling attracted some attention.
A careful analysis of the new models reveals that this quantum symmetry
exists only on one-dimensional lattices and in an appropriate approximation
seems still to be restricted to
half-filling. Despite these shortcomings 
the existence of novel symmetries in
Hubbard models is very interesting and worth investigating.
Quantum symmetries of the Hubbard model were first investigated in the form of
Yangians \cite{UK}; quantum supersymmetries have
also been considered \cite{GHLZ}.

In this letter we shall investigate the origin of quantum symmetries in
extended Hubbard models. We will find a one-to-one correspondence 
between Hamiltonians with quantum and classical symmetries.
Guided by our results we will 
then be able to identify models whose
symmetries are neither restricted to one-dimensional lattices nor to half
filling.

Originally introduced as a simplistic description of narrow d-bands in
transition metals, the Hubbard model combines band-like and atomic behavior.
In the standard Hubbard Hamiltonian
\beq
\hhub =  u \sum_i n_{i \1} n_{i \0} - \mu \sum_{i,\s} n_{i \s}
+ t \sum_{\la i,j \ra \s} a^\dag_{j \s } a_{i \s},\label{hhub}
\eeq
this is achieved by a local Coulomb term and a competing 
non-local hopping term.
Here $a^\dag_{i \s }$, $a_{i \s}$ are creation and annihilation
operators\footnote{We will use the convention that
operators at different sites {\em commute}. On a bipartite lattice
one can easily switch to anticommutators without
changing any of our results.} for electrons of
spin $\s\in\{\1, \0\}$ at site $i$ of a $D$-dimensional lattice, 
$\la i,j \ra$ denotes nearest neighbor sites and
$n_{i \s} \equiv a^\dag_{i \s }a_{i \s}$.
The average number of electrons $\la \sum_{i,\s} n_{i \s} \ra$
is fixed by the chemical potential $\mu$.
 
The {\em standard Hubbard model\/} has a 
SO(4) symmetry
at $\mu = u/2$, the value of $\mu$ corresponding 
to half filling in the
band-like limit. This symmetry is the product of a {\em magnetic\/}
$\su2_m$ (spin) with local generators
\beq
\xp_m = a^\dag_\1 a_\0,\quad \xm_m = a^\dag_\0 a_\1, 
\quad H_m = n_\1 - n_\0 ,
\label{mag}
\eeq
and a {\em superconducting\/} $\su2_s$
(pseudo-spin) with local generators
\beq
\xp_s = a^\dag_\1 a^\dag_\0, \quad \xm_s = a_\0 a_\1,
\quad H_s = n_\1 + n_\0 - 1 ,
\label{suco}
\eeq
modulo a $\zet_2$, generated by the unitary transformation
($a_\0 \leftrightarrow a^\dag_\0$) that interchanges the
two sets of local generators.
The mutually orthogonal
algebras generated by (\ref{mag}) and (\ref{suco})
are isomorphic to the algebra generated by the Pauli matrices
and have unit elements
$1_s = H_s^2$, $1_m = H_m^2$ with $1_s + 1_m = 1$.
The superconducting generators commute
with each term of the local part $\hloc$ (first two terms) 
of the Hubbard Hamiltonian (\ref{hhub})
provided that
$\mu = u/2$.
This can either be seen by direct 
computation or by studying the action of the
generators on the four possible electron states at each site.
It is also easily seen that the magnetic generators
commute with each term of $\hloc$; in the following
we will however
focus predominantly on the superconducting symmetry.

To check the symmetry of the non-local hopping term
we have to consider global generators ${\cal O}$:
These generators are here simply
given by the sum $\sum {\cal O}_i$ of the 
local generators for all sites $i$.
The rule that governs the combination of
representations for more than one lattice site is abstractly given by
the diagonal map
or coproduct $\Delta$ of $U(su(2))$.
Generators for two sites are directly obtained from the coproduct
$$
\Delta(X^\pm)  =  X^\pm \otimes 1 + 1 \otimes X^\pm, \quad
\Delta(H)  =  H \otimes 1 + 1 \otimes H,
$$
while generators for $N$ sites require 
$(N-1)$-fold iterative application of
$\Delta$. Coassociativity of $\Delta$ ensures that it does not matter which
tensor factor is split up at each step.
Another distinguishing property of this {\em classical\/} coproduct is its
symmetry (cocommutativity). This property and coassociativity ensure that we
can arrange that the two factors of the last coproduct coincide with any
given pair $\la i,j \ra$ of next-neighbor sites; see Fig.~\ref{coass}. 
It is hence enough to study
symmetry of a single next-neighbor term of the Hamiltonian
to prove global symmetry.
\begin{figure}[tbp]
\begin{center}
\unitlength 1.00mm
\linethickness{0.4pt}
\begin{picture}(71.00,9.00)
\put(0.00,4.00){\circle*{2.00}}
\put(20.00,4.00){\circle*{2.00}}
\put(30.00,4.00){\circle*{2.00}}
\put(40.00,4.00){\circle*{2.00}}
\put(50.00,4.00){\circle*{2.00}}
\put(70.00,4.00){\circle*{2.00}}
\put(60.00,4.00){\makebox(0,0)[cc]{$\cdots$}}
\put(10.00,4.00){\makebox(0,0)[cc]{$\cdots$}}
\put(0.00,9.00){\makebox(0,0)[cc]{$(\id$}}
\put(70.00,9.00){\makebox(0,0)[cc]{$\id)$}}
\put(10.00,9.00){\makebox(0,0)[cc]{$\cdots$}}
\put(60.00,9.00){\makebox(0,0)[cc]{$\cdots$}}
\put(5.00,9.00){\makebox(0,0)[cc]{$\otimes$}}
\put(15.00,9.00){\makebox(0,0)[cc]{$\otimes$}}
\put(25.00,9.00){\makebox(0,0)[cc]{$\otimes$}}
\put(45.00,9.00){\makebox(0,0)[cc]{$\otimes$}}
\put(55.00,9.00){\makebox(0,0)[cc]{$\otimes$}}
\put(65.00,9.00){\makebox(0,0)[cc]{$\otimes$}}
\put(20.00,9.00){\makebox(0,0)[cc]{$\id$}}
\put(50.00,9.00){\makebox(0,0)[cc]{$\id$}}
\put(35.00,9.00){\makebox(0,0)[cc]{$\Delta$}}
\put(30.00,4.00){\line(1,0){10.00}}
\put(30.00,1.00){\makebox(0,0)[cc]{$i$}}
\put(40.00,1.00){\makebox(0,0)[cc]{$j$}}
\end{picture}
\end{center}
\caption{Coassociativity of $\Delta$ reduces global symmetry to symmetry of
next-neighbor terms $\la i,j \ra$ if $D = 1$.} 
\label{coass}
\end{figure}

The search for quantum group symmetries in the Hubbard model is motivated by
the observation that the local generators $\xp_s$, $\xm_s$ and $H_s$ in the
superconducting representation of $\su2$ also
satisfy the $\suq2$ algebra as given in the Jimbo-Drinfel'd basis \cite{J}
\beq
\left[\xp , \xm\right] = {q^H - q^{-H} \over q - q^{-1}}, \qquad \left[H ,
X^\pm\right] = \pm 2 X^\pm.
\eeq
(The proof uses $H_s^3 = H_s$.)
It immediately follows that $\hloc$ has a local quantum symmetry.
As is, this is a trivial
statement because we did not yet consider global quantum symmetries.
Global generators are now defined via the deformed coproduct of $\suq2$,
$q \in {\bf R}\backslash\{0\}$
\bea
\Delta_q(X^\pm) & = & X^\pm \ot q^{-H/2} + q^{H/2} \ot X^\pm ,\nonumber\\
\Delta_q(H) & = & H \ot 1 + 1 \ot H.
\eea
The local symmetry can be extended to a non-trivial global quantum
symmetry by a modification of the Hubbard Hamiltonian. The idea of
\cite{MR} was to achieve this by including phonons.
Before we proceed to study the resulting extended Hubbard Hamiltonian 
$\hext$,
we would like to make two remarks: (i) We call a Hamiltonian quantum symmetric
if it commutes with all global generators. This implies invariance under the
quantum adjoint action and vice versa.
(ii) Coproducts of quantum groups are coassociative but not cocommutative. This
means that the reduction of global symmetry to that of next-neighbor terms
holds only for one-dimensional lattices. The practical implication is an
absence of quantum symmetries for higher-dimensional lattices.
(For a triangular lattice this is illustrated in Fig.~\ref{tri}.)
\begin{figure}[tbp]
\unitlength 0.75mm
\linethickness{0.4pt}
\begin{picture}(65.00,13.00)
\put(15.00,5.00){\circle*{2.00}}
\put(5.00,5.00){\circle*{2.00}}
\put(10.00,12.00){\circle*{2.00}}
\put(25.00,5.00){\circle*{2.00}}
\put(35.00,5.00){\circle*{2.00}}
\put(45.00,5.00){\circle*{2.00}}
\put(55.00,5.00){\circle*{2.00}}
\put(50.00,12.00){\circle*{2.00}}
\put(30.00,12.00){\circle*{2.00}}
\put(5.00,5.00){\line(1,0){10.00}}
\put(35.00,5.00){\line(-3,4){5.25}}
\put(45.00,5.00){\line(3,4){5.25}}
\put(20.00,8.00){\makebox(0,0)[cc]{$\sim$}}
\put(40.00,8.00){\makebox(0,0)[cc]{$\sim$}}
\put(60.00,8.00){\makebox(0,0)[cc]{$\Leftrightarrow$}}
\put(65.00,8.00){\makebox(0,0)[lc]{$\Delta$ is cocommutative}}
\end{picture}
\caption{In $D \neq 1$ symmetry of next-neighbor terms
implies global symmetry only if
$\Delta$ is classical.} 
\label{tri}
\end{figure}

The {\em extended Hubbard model} of~\cite{MR} 
(with some modifications \cite{CS}) 
introduces Einstein oscillators (parameters: $M$,
$\omega$) and electron-phonon couplings (local: $\vec\lambda$-term,
non-local: via $T_{ij\s}$):
\bea
\hext & = & u \sum_i n_{i \1} n_{i \0} - \mu \sum_{i,\s} n_{i \s}
            - \vec\lambda\cdot\sum_{i \s} n_{i \s} \vec x_i \nonumber\\
	 && + \sum_i\left({\vec {p_i}^2 \over 2 M} + {1 \over 2} M \omega^2 
	      \vec x_i^2\right) 
            + \sum_{\la i,j \ra \s} a^\dag_{j \s } a_{i \s} T_{ij\sigma},
\eea
with hopping amplitude
\beq
T_{ij\sigma} = T_{ji\sigma}^\dag = 
t \exp(\zeta \hat e_{ij}\cdot(\vec x_i - \vec x_j) 
+ i \vec\kappa\cdot(\vec p_i - \vec p_j)).
\eeq
The displacements $\vec x_i$ of the ions from their rest positions
and the corresponding momenta $\vec p_i$ satisfy
canonical commutation relations. The
$\hat e_{ij}$ are unit vectors from site $i$ to site $j$.
For $\vec\kappa = 0$ the model reduces to the Hubbard model with
phonons and atomic orbitals
$\psi(r) \sim \exp(-\zeta r)$
in $s$-wave approximation \cite{CS}.

The local part of $\hext$
commutes with the generators of $\suq2_s$ 
iff
\beq 
\mu = {u \over 2} - {\vec\lambda^2 \over M \omega^2}.
\eeq
(For technical reasons one needs to use modified generators
$\tilde X^\pm_s \equiv \exp(\mp\frac{2 i \vec\lambda\cdot\vec p}{\hbar M
\omega^2}) X^\pm_s$ here
that however still satisfy the $\suq2$ algebra.)\\
The nonlocal part of $\hext$ and thereby the whole extended
Hubbard Hamiltonian commutes with the global generators
iff
\beq
\vec\lambda = \hbar M \omega^2 \vec\kappa , \qquad
q = \exp(2 \kappa \zeta \hbar),
\eeq
where $\kappa \equiv - \hat e_{ij}\cdot\vec\kappa$
for $i,j$ ordered next neighbour sites. 
\emph{For $q \neq 1$ the symmetry is restricted to models
given on a 1-dimensional lattice with naturally ordered sites.}

From what we have seen so far we could be let to the premature conclusion
that the quantum symmetry is due to phonons and that we have found symmetry
away from half filling because $\mu \neq u/2$. However: the pure Hubbard
model with phonons has $\vec\kappa = 0$ and hence a classical symmetry $(q=1)$.
Furthermore $\vec\lambda \neq 0$ implies non-vanishing local electron-phonon
coupling so that a mean field approximation cannot be performed and we simply
do not know how to compute the actual filling.
Luckily there is an equivalent model that is not plagued with this problem:
A {\em Lang-Firsov transformation} 
with unitary operator $U = \exp(i \vec\kappa\cdot
\sum_j \vec p_j n_{j \s})$.
leads to the Hamiltonian 
\beq
\hqsym = U \hext U^{-1} = \hext(\vec\lambda',u',\mu',{T'}_{ij\sigma}),
\label{LF}
\eeq
of what we shall call the {\em quantum symmetric
Hubbard model}. It has the same form as $\hext$, but
with a new set of parameters 
\bea
\vec\lambda' & = & \vec\lambda - M \omega^2 \hbar \vec\kappa\\
u' & = & u - 2 \hbar \vec\lambda\cdot\vec\kappa + M \omega^2 \hbar^2 \kappa^2\\ 
\mu' & = & \mu + \hbar \vec\lambda\cdot\vec\kappa 
- {1 \over 2}M \omega^2 \hbar^2 \kappa^2
\eea 
and a modified hopping amplitude
\beq
T'_{ij,-\s} = \tilde t_{ij} (1 + 
(q^{{\hat e_{ji} \over 2}} - 1) n_{i \s})(1 + 
(q^{{\hat e_{ij} \over 2}} -1) n_{j \s})
\eeq
where $\tilde t_{ij} = t \exp(\zeta \hat e_{ij}\cdot (\vec x_i - \vec x_j))$. 
The condition for symmetry expressed in terms of the new parameters is 
\beq
\vec\lambda' =0,\qquad \mu' = {u' \over 2},
\eeq
\ie requires vanishing local phonon coupling
and corresponds to {\em half filling}!
$\tilde t_{ij}$ may also be turned
into a (temperature-dependent) constant via a mean field approximation. This
approximation is admissible for the quantum symmetric Hubbard model because
$\vec\lambda' = 0$.

\begin{figure}[htb]
\begin{center}
\unitlength 3.50pt
\linethickness{0.4pt}
\begin{picture}(61.00,56.00)
\put(26.00,26.00){\circle*{2.00}}
\put(27.00,27.00){\line(1,1){3.00}}
\put(31.00,31.00){\circle{2.00}}
\put(32.00,32.00){\line(1,1){3.00}}
\put(36.00,36.00){\circle*{2.00}}
\put(27.00,26.00){\line(1,0){5.00}}
\put(33.00,26.00){\circle{2.00}}
\put(34.00,26.00){\line(1,0){5.00}}
\put(40.00,26.00){\circle*{2.00}}
\put(41.00,27.00){\line(1,1){3.00}}
\put(45.00,31.00){\circle{2.00}}
\put(37.00,36.00){\line(1,0){5.00}}
\put(43.00,36.00){\circle{2.00}}
\put(46.00,32.00){\line(1,1){3.00}}
\put(50.00,36.00){\circle*{2.00}}
\put(44.00,36.00){\line(1,0){5.00}}
\put(12.00,26.00){\circle*{2.00}}
\put(13.00,27.00){\line(1,1){3.00}}
\put(17.00,31.00){\circle{2.00}}
\put(18.00,32.00){\line(1,1){3.00}}
\put(22.00,36.00){\circle*{2.00}}
\put(13.00,26.00){\line(1,0){5.00}}
\put(19.00,26.00){\circle{2.00}}
\put(20.00,26.00){\line(1,0){5.00}}
\put(23.00,36.00){\line(1,0){5.00}}
\put(29.00,36.00){\circle{2.00}}
\put(30.00,36.00){\line(1,0){5.00}}
\put(37.00,37.00){\line(1,1){3.00}}
\put(41.00,41.00){\circle{2.00}}
\put(42.00,42.00){\line(1,1){3.00}}
\put(46.00,46.00){\circle*{2.00}}
\put(51.00,37.00){\line(1,1){3.00}}
\put(55.00,41.00){\circle{2.00}}
\put(47.00,46.00){\line(1,0){5.00}}
\put(53.00,46.00){\circle{2.00}}
\put(56.00,42.00){\line(1,1){3.00}}
\put(60.00,46.00){\circle*{2.00}}
\put(54.00,46.00){\line(1,0){5.00}}
\put(23.00,37.00){\line(1,1){3.00}}
\put(27.00,41.00){\circle{2.00}}
\put(28.00,42.00){\line(1,1){3.00}}
\put(32.00,46.00){\circle*{2.00}}
\put(33.00,46.00){\line(1,0){5.00}}
\put(39.00,46.00){\circle{2.00}}
\put(40.00,46.00){\line(1,0){5.00}}
\put(36.00,37.00){\line(0,1){5.00}}
\put(36.00,43.00){\circle{2.00}}
\put(36.00,35.00){\line(0,-1){5.00}}
\put(36.00,29.00){\circle{2.00}}
\put(29.41,34.97){\line(2,-5){1.18}}
\put(29.63,36.86){\line(1,1){5.53}}
\put(29.99,36.39){\line(5,2){10.07}}
\put(40.06,40.42){\line(0,0){0.00}}
\put(36.97,42.60){\line(5,-2){3.05}}
\put(36.79,42.31){\line(1,-1){5.53}}
\put(41.62,40.13){\line(1,-3){1.05}}
\put(32.03,31.19){\line(5,2){10.18}}
\put(29.85,35.37){\line(1,-1){5.53}}
\put(40.60,40.06){\line(-2,-5){4.06}}
\put(35.50,42.07){\line(-2,-5){4.04}}
\put(31.92,30.49){\line(5,-2){3.08}}
\put(36.87,29.59){\line(1,1){5.57}}
\put(36.00,44.00){\line(0,1){5.00}}
\put(36.00,28.00){\line(0,-1){5.00}}
\put(36.00,22.00){\circle*{2.83}}
\put(36.00,50.00){\circle*{2.83}}
\put(36.00,51.00){\line(0,1){5.00}}
\put(36.00,21.00){\line(0,-1){5.00}}
\put(36.00,16.00){\line(0,0){0.00}}
\put(36.00,16.00){\line(0,0){0.00}}
\put(36.00,16.00){\line(0,0){0.00}}
\put(24.00,28.00){\line(0,1){6.00}}
\put(24.00,27.00){\circle*{2.83}}
\put(24.00,35.00){\circle*{2.83}}
\put(24.00,36.00){\line(0,1){5.00}}
\put(24.00,26.00){\line(0,-1){5.00}}
\put(14.00,3.00){\circle*{2.00}}
\put(15.00,4.00){\line(1,1){3.00}}
\put(19.00,8.00){\circle{2.00}}
\put(20.00,9.00){\line(1,1){3.00}}
\put(24.00,13.00){\circle*{2.00}}
\put(15.00,3.00){\line(1,0){5.00}}
\put(21.00,3.00){\circle{2.00}}
\put(22.00,3.00){\line(1,0){5.00}}
\put(28.00,3.00){\circle*{2.00}}
\put(29.00,4.00){\line(1,1){3.00}}
\put(33.00,8.00){\circle{2.00}}
\put(25.00,13.00){\line(1,0){5.00}}
\put(31.00,13.00){\circle{2.00}}
\put(34.00,9.00){\line(1,1){3.00}}
\put(38.00,13.00){\circle*{2.00}}
\put(32.00,13.00){\line(1,0){5.00}}
\put(0.00,3.00){\circle*{2.00}}
\put(1.00,4.00){\line(1,1){3.00}}
\put(5.00,8.00){\circle{2.00}}
\put(6.00,9.00){\line(1,1){3.00}}
\put(10.00,13.00){\circle*{2.00}}
\put(1.00,3.00){\line(1,0){5.00}}
\put(7.00,3.00){\circle{2.00}}
\put(8.00,3.00){\line(1,0){5.00}}
\put(11.00,13.00){\line(1,0){5.00}}
\put(17.00,13.00){\circle{2.00}}
\put(18.00,13.00){\line(1,0){5.00}}
\put(25.00,14.00){\line(1,1){3.00}}
\put(29.00,18.00){\circle{2.00}}
\put(30.00,19.00){\line(1,1){3.00}}
\put(34.00,23.00){\circle*{2.00}}
\put(39.00,14.00){\line(1,1){3.00}}
\put(43.00,18.00){\circle{2.00}}
\put(35.00,23.00){\line(1,0){5.00}}
\put(41.00,23.00){\circle{2.00}}
\put(44.00,19.00){\line(1,1){3.00}}
\put(48.00,23.00){\circle*{2.00}}
\put(42.00,23.00){\line(1,0){5.00}}
\put(11.00,14.00){\line(1,1){3.00}}
\put(15.00,18.00){\circle{2.00}}
\put(16.00,19.00){\line(1,1){3.00}}
\put(20.00,23.00){\circle*{2.00}}
\put(21.00,23.00){\line(1,0){5.00}}
\put(27.00,23.00){\circle{2.00}}
\put(28.00,23.00){\line(1,0){5.00}}
\put(24.00,14.00){\line(0,1){5.00}}
\put(24.00,20.00){\circle{2.00}}
\put(24.00,12.00){\line(0,-1){5.00}}
\put(24.00,6.00){\circle{2.00}}
\put(17.41,11.97){\line(2,-5){1.18}}
\put(17.63,13.86){\line(1,1){5.53}}
\put(17.99,13.39){\line(5,2){10.07}}
\put(28.06,17.42){\line(0,0){0.00}}
\put(24.97,19.60){\line(5,-2){3.05}}
\put(24.79,19.31){\line(1,-1){5.53}}
\put(29.62,17.13){\line(1,-3){1.05}}
\put(20.03,8.19){\line(5,2){10.18}}
\put(17.85,12.37){\line(1,-1){5.53}}
\put(28.60,17.06){\line(-2,-5){4.06}}
\put(23.50,19.07){\line(-2,-5){4.04}}
\put(19.92,7.49){\line(5,-2){3.08}}
\put(24.87,6.59){\line(1,1){5.57}}
\put(24.00,5.00){\line(0,-1){5.00}}
\put(35.00,-5.00){\line(0,0){0.00}}
\put(35.00,-5.00){\line(0,0){0.00}}
\put(35.00,-5.00){\line(0,0){0.00}}
\put(36.00,15.00){\circle*{2.83}}
\put(36.00,14.00){\line(0,-1){5.00}}
\end{picture}
\end{center}
\caption{Typical cuprate superconductor 
with CuO${}_2$ conduction planes.} \label{cuprate}
\end{figure}
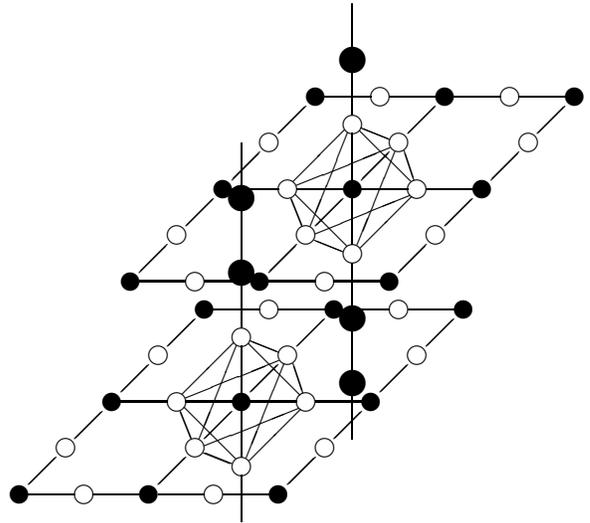

We have so far identified several quantum group symmetric models (with and
without phonons) and have achieved a better understanding of $\hext$'s
superconducting quantum symmetry. There are however still open questions:
(i) Does a new model exist that is equivalent to $\hqsym$ in 1-D but can
also be formulated on higher dimensional lattices without breaking the
symmetry? This would be important for realistic models, see
Fig.~\ref{cuprate}.
(ii) Are there models with symmetry away from half-filling?
(iii) What is the precise relation between models with
classical and quantum symmetry in
this setting?

As we shall see the answer to the last question also leads to the resolution
of the first two. Without loss of generality (see argument given above)
we will focus on one pair of next-neighbor sites in the following.
We shall present two approaches that supplement each other:

\paragraph{Generalized Lang-Firsov transformation}
We recall that the Hubbard model with phonons (with classical symmetry)
can be transformed into the standard Hubbard
Hamiltonian in two steps: A Lang-Firsov transformation changes the
model to one with vanishing local phonon coupling and a mean field
approximation removes the phonon operators from the model by averaging
over Einstein oscillator eigenstates \cite{RMR}. There
exists a similar transformation that relates the extended Hubbard model
(with quantum symmetry) to the standard Hubbard model:
$$
\hext \longleftrightarrow \hqsym \longleftrightarrow \hhub.
$$
(We have already
seen the first step of this transformation above in (\ref{LF}).)
It is easy to see that the hopping terms of $\hqsym$ and $\hhub$ have
different spectrum so the transformation that we are looking for
cannot be an equivalence
transformation. There exists however an invertible operator $M$, with 
$M M^* = 1 +  (\alpha^2 -1) \xi$, 
$\xi^2 = \xi$ (\ie similar to a partial isometry), that transforms the
coproducts of the classical 
Chevalley generators into their
Jimbo-Drinfel'd quantum counterparts
\bea
M \Delta_c(X^\pm)_s M^* & = & \Delta_q(X^\pm)_s \nonumber \\
M \Delta_c(H)_s M^* & = & \Delta_q(H)_s, \label{copxh}
\eea
and the standard Hubbard Hamiltonian into $\hqsym$
\beq
M  \hhub M^* = \hqsym  .
\eeq
This operator $M$ is
\beq
M = 1 \ot 1 + (\alpha -1) \xi + \beta f,
\eeq
with $f = X^-_s \ot X^+_s - X^+_s \ot X^-_s$,
$\xi = - f^2 = {1 \over 2}(H_s^2 \ot H_s^2 - H_s \ot H_s)$
and $\alpha \pm \beta = q^{\pm {1 \over 2}}$.
With this knowledge the proof of the quantum symmetry of $\hext$ is greatly
simplified.

\paragraph{Quantum vs.\ Classical Groups---Twists}
A systematic way to study the relation of quantum and classical symmetries
was given by Drinfel'd \cite{D}. 
He argues that the classical U(g) and $q$-deformed
U$_q$(g) universal enveloping algebras are isomorphic {\em as algebras}.
The relation of the Hopf algebra structures is slightly more involved:
the undeformed universal enveloping algebra U(g) of a
Lie algebra, interpreted as a quasi-associative Hopf
algebra whose coassociator is an invariant 3-tensor,
is twist-equivalent to the Hopf algebra U$_q$(g) (over $[[\ln q]]$).

All we need to know here is that
classical ($\Delta_c$) and quantum ($\Delta_q$)
coproducts are related via conjugation (``twist'')
by the so-called 
universal ${\cal F} \in \mbox{U}_q(\mbox{su}_2)^{\hat\ot 2}$:
\beq
\Delta_q (x) = {\cal F} \Delta_c(x) {\cal F}^{-1}. 
\label{copt}
\eeq
(For notational simplicity we did not explicitly write the map that describes
the algebra isomorphism of $\mbox{U}(\mbox{su}_2)$ and 
$\mbox{U}_q(\mbox{su}_2)$ but
we should not be fooled by the apparent similarity between
(\ref{copxh}) and (\ref{copt}): The algebra isomorphism does not
map Chevalley generators to Jimbo-Drinfel'd generators  
and $M$ is not a representation of ${\cal F}$.)

The fundamental matrix representation of the universal ${\cal F}$
for SU(N) is an orthogonal matrix \cite{E}
\bea
\rho^{\ot 2}({\cal F}) 
& = & \sum_i e_{ii} \ot e_{ii} 
      + \cos\varphi \sum_{i \neq j} e_{ii} \ot e_{jj} \nonumber \\
& + & \sin\varphi \sum_{i < j} \left( e_{ij} \ot e_{ji} - e_{ji} \ot
      e_{ij}\right),
\eea
where $\cos\varphi \pm \sin\varphi = \sqrt{2 q^{\pm 1}/(q + q^{-1})}$ ,
$i,j = 1\ldots N$ and $e_{ij}$ are N$\times$N matrices with lone
``1'' at position $(i,j)$.
The universal ${\cal F}$ in the superconducting
spin-${1\over 2}$ representation, {\em i.e.} essentially
the $N = 2$ case with the Pauli
matrices replaced by (\ref{suco}), is
\beq
F_s = \exp(\varphi f)_s = \tilde\xi + \cos\varphi \,\xi 
      + \sin\varphi \,f
\eeq
and $\tilde\xi + \xi = 1_s \ot 1_s$.
We are interested in a representation of the universal
${\cal F}$ on the 16-dimensional Hilbert
space of states of two sites:
\bea
F & = & (\epsilon_m \oplus \rho_s)^{\otimes 2}({\cal F}) 
        = \exp(\varphi f) \nonumber\\
  & = & 1 \ot 1 - 1_s \ot 1_s + F_s.
\eea
Note that the trivial magnetic representation $\epsilon_m$ enters here
even though we decided to study only deformations of the superconducting
symmetry---$F_s$ alone would have been
identically zero on the hopping term
and would hence have lead to a trivial model.

We now face a puzzle: By construction $F^{-1} \hqsym F$ should
commute with the (global)
generators of $\suq2_s$ just like $\hhub$.
But $F^{-1} \hqsym F$ obviously has the same spectrum
as $\hqsym$ so it cannot be equal to $\hhub$.
There must be other models with the same symmetries.
In fact we find a six-parameter family of classically 
symmetric models in any dimension.
In the one-dimensional case twist-equivalent quantum symmetric models
can be constructed as deformations of each of these classical models.
$\hhub$ and $\hqsym$ are not a twist-equivalent pair but
all models mentioned are related by 
generalized Lang-Firsov transformations.

To close we would like to present the most general Hamiltonian
with $\su2\times\su2/\zet_2$ symmetry and symmetric next-neighbor
terms.
(A group-theoretical derivation and detailed description of this
model is however beyond the scope of this letter and will be given elsewhere.)
The Hamiltonian is written with eight real 
parameters ($\mu$, $r$, $s$, $t$, $u$, $v$, Re($z$), Im($z$)):
\bea
\hsym \:
&& =  u \sum_i n_{i\1} n_{i\0} \quad - \: \mu \sum_{i,\s} n_{i\s}
      \quad + \: t \sum_{\la i,j \ra \s} a^\dag_{i\s} a_{j\s}\nonumber\\
&& +  \: r \sum_{\la i,j \ra \s} n_{i\s} n_{j -\s}
      \quad +  \: s \sum_{\la i,j \ra \s} n_{i\s} n_{j \s} \nonumber\\    
&& +  \: {2 \mu - u \over e} \sum_{\la i,j \ra \s}
      a^\dag_{i\1} a^\dag_{i\0} a_{j\1} a_{j\0} \nonumber\\ 
&& +  \: (s - r) \sum_{\la i,j \ra \s} a^\dag_{i\s} a_{i -\s} a^\dag_{j -\s}
      a_{j \s} \nonumber\\
&& +  \: v \sum_{\la i,j \ra \s} 
      \left(n_{i\1} n_{i\0} n_{j\1} n_{j\0}
      - n_{i\1} n_{i\0} n_{j\s} 
      - n_{i\s} n_{j\1} n_{j\0}\right) \nonumber\\ 
&& +  \sum_{\la i,j \ra \s} a^\dag_{i -\s} a_{j -\s}
      \left(z(n_{i\s} - 1) n_{j\s} + z^* n_{i\s} (n_{j\s} -1)\right) 
      \nonumber\\    
&&    \mbox{\sc + h.c.} 
\eea
For symmetry $v = r + s + u - 2 \mu$ must hold.
One parameter can be absorbed into an overall multiplicative
constant, so we have six free parameters.
The first three terms comprise the standard Hubbard model but now
without the restriction to half-filling.
The filling factor is fixed by the coefficient of the
pair hopping term (6$^{th}$ term). The number $e$ in
the denominator of this coefficient 
is the number of edges per site. For a single pair of
sites $e = 1$, for a one-dimensional chain $e = 2$, for a honeycomb lattice
$e = 3$, for a square lattice $e = 4$, for a triangular lattice $e = 6$
and for a $D$-dimensional hypercube $e = 2 D$.
For a model on a general graph $e$ will vary with the site.
The 4$^{th}$ and 5$^{th}$ term describe density-density interaction 
for anti-parallel and parallel spins respectively.
The balance of these two interactions is governed by the
coefficient of the spin-wave term (7$^{th}$ term).
The last term is a modified hopping term that is reminiscent of the
hopping term in the $t$-$J$ model with hopping strength depending
on the occupation of the sites;
after deformation this term is the origin of the 
non-trivial quantum symmetries
of $\hsym$.

The known and many new quantum symmetric Hubbard models can be derived 
from $\hsym$ by twisting as described above.
While the deformation provides up to two extra parameters for the
quantum symmetric models the 
advantage of the corresponding classical
models is that they are not restricted to one dimension.
There are both classically and quantum symmetric
models with symmetries away from half filling.

The way the filling and the spin-spin interactions appear as coefficients
of the pair-hopping and spin-wave terms respectively looks quite
promising for a physical interpretation.
Due to its symmetries $\hsym$ should share some of the
nice analytical properties of the standard Hubbard Hamiltonian and
could hence be of interest in its own right.

\section*{Acknowledgments}

It is a pleasure to thank Julius Wess for initiating this line of research and
Bianca L.\ Cerchiai for early collaboration.
Very helpful conversations with Bruno Zumino, Nicolai Reshetikhin, Michael
Schlieker, Chong-Sun Chu, Pei-Ming Ho, Harold Steinacker and Bogdan Morariu
at UC Berkeley and Orlando Alvarez, Paul Watts and their colleagues 
at the University of Miami are gratefully acknowledged. I would also like to
thank Elliott Lieb for pointing out the relevance of the hidden symmetries.

This work was supported in part by 
the Director, Office of Energy Research,
Office of High Energy and Nuclear Physics, Division of High Energy Physics of
the U.S.\ Department of Energy under Contract DE-AC03-76SF00098, in part by
the National Science Foundation under grant PHY-90-21139 and in part by the
Max Planck Institut f\"ur Physik in Munich.

\end{document}